\documentclass[12pt]{article}
\usepackage{amssymb}
\begin{document}

\newcommand{\sect}[1]{\setcounter{equation}{0}\section{#1}}
\renewcommand{\theequation}{\thesection.\arabic{equation}}
\newcommand{\be}{\begin{equation}}
\newcommand{\ee}{\end{equation}}
\newcommand{\bea}{\begin{eqnarray}}
\newcommand{\eea}{\end{eqnarray}}
\newcommand{\nonu}{\nonumber\\}
\newcommand{\beano}{\begin{eqnarray*}}
\newcommand{\eeano}{\end{eqnarray*}}
\newcommand{\eps}{\epsilon}
\newcommand{\om}{\omega}
\newcommand{\vph}{\varphi}
\newcommand{\sig}{\sigma}
\newcommand{\CC}{\mbox{${\mathbb C}$}}
\newcommand{\RR}{\mbox{${\mathbb R}$}}
\newcommand{\QQ}{\mbox{${\mathbb Q}$}}
\newcommand{\ZZ}{\mbox{${\mathbb Z}$}}
\newcommand{\NN}{\mbox{${\mathbb N}$}}
\newcommand{\1}{\mbox{\hspace{.0em}1\hspace{-.24em}I}}
\newcommand{\II}{\mbox{${\mathbb I}$}}
\newcommand{\prt}{\partial}
\newcommand{\und}[1]{\underline{#1}}
\newcommand{\wh}[1]{\widehat{#1}}
\newcommand{\wt}[1]{\widetilde{#1}}
\newcommand{\mb}[1]{\ \mbox{\ #1\ }\ }
\newcommand{\half}{\frac{1}{2}}
\newcommand{\noin}{\not\!\in}
\newcommand{\rhotimes}{\mbox{\raisebox{-1.2ex}{$\stackrel{\displaystyle\otimes}
{\mbox{\scriptsize{$\rho$}}}$}}}
\newcommand{\bin}[2]{{\left( {#1 \atop #2} \right)}}
\newcommand{\A}{\cal A}
\newcommand{\B}{\cal B}
\newcommand{\C}{\cal C}
\newcommand{\F}{{\cal F}}
\newcommand{\R}{{\cal R}}
\newcommand{\T}{{\cal T}}
\newcommand{\cS}{{\cal S}}
\newcommand{\hlp}{{\RR}_+}
\newcommand{\hlm}{{\RR}_-}
\newcommand{\Hil}{{\cal H}}
\newcommand{\D}{{\cal D}}
\newcommand{\brep}{{\cal F}_B}
\newcommand{\form}{\langle \, \cdot \, , \, \cdot \, \rangle }
\newcommand{\e}{{\rm e}}
\newcommand{\Rp}{{R^+_{\, \, \, \, }}}
\newcommand{\Rm}{{R^-_{\, \, \, \, }}}
\newcommand{\Rpm}{{R^\pm_{\, \, \, \, }}}
\newcommand{\Tp}{{T^+_{\, \, \, \, }}}
\newcommand{\Tm}{{T^-_{\, \, \, \, }}}
\newcommand{\Tpm}{{T^\pm_{\, \, \, \, }}}
\newcommand{\baral}{\bar{\alpha}}
\newcommand{\barbt}{\bar{\beta}}
\newcommand{\EE}{\mbox{${\mathbb E}$}}
\newcommand{\JJ}{\mbox{${\mathbb J}$}}
\newcommand{\MM}{\mbox{${\mathbb M}$}}
\newcommand{\ct}{{\cal T}}

\newtheorem{theo}{Theorem}[section]
\newtheorem{coro}[theo]{Corollary}
\newtheorem{prop}[theo]{Property}
\newtheorem{defi}[theo]{Definition}
\newtheorem{conj}[theo]{Conjecture}
\newtheorem{lem}[theo]{Lemma}
\newcommand{\prf}{\underline{Proof:}\ }
\newcommand{\finprf}{\null \hfill {\rule{5pt}{5pt}}\\[2.1ex]\indent}

\pagestyle{empty}
\rightline{August 2002}

\vfill

\begin{center}
{\Large\bf Scattering in the Presence of a Reflecting \\[1.2ex]
and Transmitting Impurity}
\\[2.1em]

{\large
M. Mintchev$^{a}$\footnote{mintchev@df.unipi.it},
E. Ragoucy$^{b}$\footnote{ragoucy@lapp.in2p3.fr}
and P. Sorba$^{b}$\footnote{sorba@lapp.in2p3.fr}}\\
\end{center}

\null

\noindent
{\it $^a$ INFN and Dipartimento di Fisica, Universit\'a di
     Pisa, Via Buonarroti 2, 56127 Pisa, Italy\\[2.1ex]
$^b$ LAPTH, 9, Chemin de Bellevue, BP 110, F-74941 Annecy-le-Vieux
     cedex, France}
\vfill

\begin{abstract}
We investigate factorized scattering from a reflecting and 
transmitting impurity. Bulk scattering
is non trivial, provided that the bulk scattering matrix depends 
separately on the spectral parameters
of the colliding particles, and not only on their difference. We show 
that a specific extension
of a boundary algebra encodes the underlying scattering theory. The 
total scattering operator
is constructed in this framework and shown to be unitary.

\end{abstract}

\vfill
\rightline{LAPTH-928/02}
\rightline{IFUP-TH 32/2002}
\rightline{\tt hep-th/0209052}
\newpage
\pagestyle{plain}
\setcounter{page}{1}

\sect{Introduction}

\null

Integrable quantum field theories with boundaries
\cite{Cherednik:vs, Sklyanin:yz, Ghoshal:tm} have been subject
of intense study during the past decade. The great interest in such 
theories stems from the
large number of potential applications in different physical areas, 
including open strings,
branes, boundary conformal field theory, dissipative quantum 
phenomena and impurity problems.
The investigations have been mainly focussed on purely reflecting boundaries.
In realistic impurity problems \cite{Saleur:1998hq, Saleur:2000gp} however,
one must often deal with defects, which both reflect and transmit.
In spite of some progress, the results
\cite{Delfino:1994nr, Konik:1997gx, Castro-Alvaredo:2002fc} on this subject
derived by the inverse scattering method, are not very
encouraging. The present status of the problem can be summarized as follows
\cite{Castro-Alvaredo:2002fc}: simultaneous reflection and 
transmission are possible only if the two-body
bulk scattering matrix is constant (momentum and energy independent).
This condition severely restricts the
class of  admissible systems. If the scattering matrix is diagonal 
for instance,
one is left \cite{Delfino:1994nr} with free bosons, free fermions and 
Federbush type models
\cite{F}, all of them representing  a limited physical interest. In 
the present paper we
explore the possibility to weaken the assumptions adopted in
\cite{Delfino:1994nr, Konik:1997gx, Castro-Alvaredo:2002fc} in order 
to avoid the above
mentioned no-go theorem, preserving at the same time the basic
physical features of integrability, reflection and transmission.
Our analysis is constructive
and uses a certain extension of the boundary algebra 
\cite{Liguori:1998xr}, previously applied
\cite{Gattobigio:1998hn, Gattobigio:1998si, Mintchev:2001aq} in the 
case of reflecting
boundaries. The framework
covers a large class of integrable systems with reflecting and
transmitting impurities and allows to
construct explicitly the total scattering operator and to prove
unitarity and asymptotic completeness.

The paper is organized as follows. The next section is devoted to the 
consistency
relations, following from three-body factorized scattering in 
presence of impurity.
In Sect. 3 we describe the general solution of these relations and 
give some examples.
An algebraic framework for deriving the scattering amplitudes is 
developed in Sect. 4.
Here we construct also the asymptotic states. In Sect. 5 we compute some
amplitudes explicitly, we define the total scattering operator and 
discuss unitarity. The last
section contains our conclusions.

\sect{Kinematics and consistency relations}

Following the basic ideas of the quantum inverse scattering method in 
1+1 dimensions,
we parameterize the asymptotic particles by their energy $E$, momentum $p$ and
an ``isotopic" index $i=1,...,N$, the latter describing the internal 
degrees of freedom.
Usually $E$ and $p$ are not independent and obey some dispersion relation.
It is conveniently implemented by parameterizing both $E$ and $p$
in terms of one parameter $\chi \in \RR$, i.e.
\be
E = E(\chi)\, , \qquad p = p(\chi)\, .
\label{gen}
\ee
The conventional relativistic dispersion relation reads
\be
E(\chi) = m \cosh (\chi)\, , \qquad p(\chi) = m \sinh (\chi)\, ,
\label{rel}
\ee
where $m$ is the mass and $\chi $ the rapidity. A non relativistic example is
\be
E(\chi) = \frac{m\chi^2}{2} + U\, , \qquad p(\chi) = m\chi \, ,
\label{nrel}
\ee
$\chi$ being the velocity and $U$ some constant. Notice that a 
Lorentz boost in (\ref{rel}) and
a Galilean transformation in (\ref{nrel}) are both realized by a translation
$\chi \mapsto \chi + \alpha$.

In what follows we adopt a generic
dispersion relation (\ref{gen}) and parameterize each asymptotic particle
by $\chi$ and its isotopic type $i$, referring to $\chi $ as spectral
parameter. We start by
considering an impurity localized at $x=0$ and without internal 
degrees of freedom. In this case
the fundamental building blocks of factorized scattering are:
\begin{description}

\item {(i)} the two-body bulk scattering matrix 
$S_{i_1i_2}^{j_1j_2}(\chi_1,\chi_2)$;

\item {(ii)} the reflection matrix  $R_i^j(\chi )$, describing
the reflection of a particle from the impurity;

\item{(iii)} the transmission matrix $T_i^j(\chi )$, describing the
transmission of a particle by the impurity.

\end{description}
It is worth stressing that
$S$ is allowed to depend on $\chi_1$ and $\chi_2$ separately \cite{Liguori:de}.
This feature represents an essential difference with respect to the 
framework of
\cite{Delfino:1994nr, Konik:1997gx, Castro-Alvaredo:2002fc}, where 
$S$ is assumed
to depend on $\chi_1-\chi_2$ only. Notice that this last condition
imposes on $S$ some symmetries; with the dispersion relation 
(\ref{rel}) for instance,
$S_{i_1i_2}^{j_1j_2}(\chi_1-\chi_2)$ turns out to be Lorentz invariant.
The same conclusion holds for Galilean invariance in the
non relativistic case with dispersion relation (\ref{nrel}).
However, one can expect in general that the effect of the impurity is not
localized only at $x=0$, but propagates also in the bulk, breaking down
the Lorentz or Galilean invariance of $S$. For this reason
we find the framework of \cite{Delfino:1994nr, Konik:1997gx, 
Castro-Alvaredo:2002fc}
too restrictive and a bit artificial at this point.
In fact, we will show below that allowing $S$ to depend on
$\chi_1$ and $\chi_2$ separately, leads to a natural generalization 
of the inverse scattering
method, which avoids the no-go theorem of
\cite{Delfino:1994nr, Konik:1997gx, Castro-Alvaredo:2002fc} and
describes a large set of integrable systems, not covered there.

It is convenient to define at this stage the matrices
\be
R^\pm (\chi) \equiv \theta(\pm \chi) R(\chi) \, ,\qquad
T^\pm (\chi) \equiv \theta(\pm \chi) T(\chi) \, ,
\label{RT}
\ee
which have a simple physical interpretation: $R^\pm$ describe the reflection
of a particle propagating $\RR_\pm$, whereas $T^\pm$ correspond to 
the transmission
of a particle from $\hlm$ to $\hlp$ and vice versa.
If one interprets the impurity as an infinitely heavy body, the data
$\{S, R^\pm, T^\pm \}$ capture all two-body interactions. They are represented
graphically in
Fig. 1.

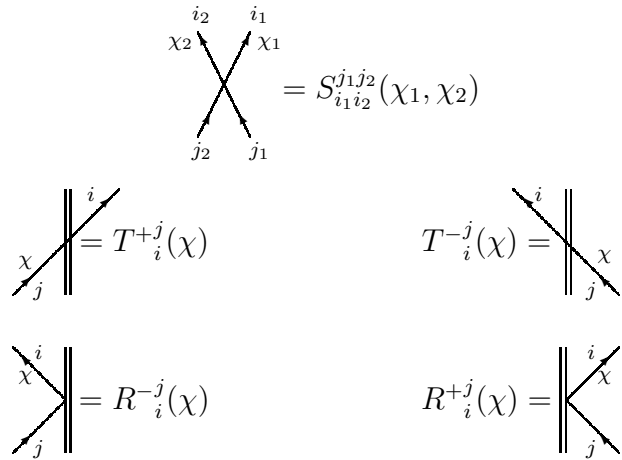
\begin{figure}[h]
\setlength{\unitlength}{0.7mm}
\begin{picture}(60,100)(-20,10)
%
%
\qbezier(65,80)(60,90)(55,100)
\qbezier(55,80)(60,90)(65,100)
\put(56.5,82.5){\vector(1,2){1}}
\put(64,97.5){\vector(1,2){1}}
\put(64,82.5){\vector(-1,2){1}}
\put(56.5,97.5){\vector(-1,2){1}}
\put(80,73){\makebox(20,20)[t]{$=S_{i_1i_2}^{j_1j_2}(\chi_1,\chi_2)$}}
\put(57,85){\makebox(20,20)[t]{${}_{i_1}$}}
\put(46,85){\makebox(20,20)[t]{${}_{i_2}$}}
\put(57,59){\makebox(20,20)[t]{${}_{j_1}$}}
\put(46,59){\makebox(20,20)[t]{${}_{j_2}$}}
\put(59,79){\makebox(20,20)[t]{${}_{\chi_1}$}}
\put(42,79){\makebox(20,20)[t]{${}_{\chi_2}$}}
%
%
\put (30,50){\line(0,0){20}}
\put (31,50){\line(0,0){20}}
\qbezier(20,50)(30,60)(40,70)
\put(22.5,52.5){\vector(1,1){1}}
\put(37.5,67.5){\vector(1,1){1}}
\put(35,44){\makebox(20,20)[t]{$=\Tp_i^j(\chi)$}}
\put(15,32.5){\makebox(20,20)[t]{${}_j$}}
\put(25,50.5){\makebox(20,20)[t]{${}_i$}}
\put(12.5,37.5){\makebox(20,20)[t]{${}_\chi$}}
%
%
\put (125,50){\line(0,0){20}}
\put (126,50){\line(0,0){20}}
\qbezier(135,50)(125,60)(115,70)
\put(132.5,52.5){\vector(-1,1){1}}
\put(118,67.5){\vector(-1,1){1}}
\put(100,44){\makebox(20,20)[t]{$\Tm_i^j(\chi)=$}}
\put(120,32.5){\makebox(20,20)[t]{${}_j$}}
\put(110,50.5){\makebox(20,20)[t]{${}_i$}}
\put(122.5,38.5){\makebox(20,20)[t]{${}_\chi$}}
%
%
%
\put (30,20){\line(0,0){20}}
\put (31,20){\line(0,0){20}}
\qbezier(20,20)(25,25)(30,30)
\qbezier(20,40)(25,35)(30,30)
\put(22.5,22.5){\vector(1,1){1}}
\put(22.5,37.5){\vector(-1,1){1}}
\put(35,14){\makebox(20,20)[t]{$=\Rm_i^j(\chi)$}}
\put(15,2.5){\makebox(20,20)[t]{${}_j$}}
\put(15,20.5){\makebox(20,20)[t]{${}_i$}}
\put(12.5,15.5){\makebox(20,20)[t]{${}_\chi$}}
%
%
\put (124,20){\line(0,0){20}}
\put (125,20){\line(0,0){20}}
\qbezier(135,20)(130,25)(125,30)
\qbezier(135,40)(130,35)(125,30)
\put(132.5,22.5){\vector(-1,1){1}}
\put(132.5,37.5){\vector(1,1){1}}
\put(100,14){\makebox(20,20)[t]{$\Rp_i^j(\chi)=$}}
\put(120,2.5){\makebox(20,20)[t]{${}_j$}}
\put(120,20.5){\makebox(20,20)[t]{${}_i$}}
\put(122.5,15.5){\makebox(20,20)[t]{${}_\chi$}}
\end{picture}
\caption{The two-body processes.}
\end{figure}

\noindent Time is flowing in these diagrams along the vertical 
direction and single lines
represent particles.
The double line corresponds to the world line of the impurity, which 
is vertical
because the impurity does not move from the point $x=0$.

The construction of the possible three-body processes
in terms of $S$, $\Rpm$ and $\Tpm$ leads to a series of relations
\cite{Delfino:1994nr, Konik:1997gx, Castro-Alvaredo:2002fc} among them.
The consistency condition, stemming from the scattering of three 
particles among
themselves, is the well known quantum Yang-Baxter equation (in its braid form)
\be
S_{12}(\chi_1,\chi_2) S_{23}(\chi_1,\chi_3) S_{12} (\chi_2,\chi_3)
= S_{23}(\chi_2,\chi_3) S_{12}(\chi_1,\chi_3) S_{23}(\chi_1,\chi_2)  \, ,
\label{qyb}
\ee
where standard tensor notation has been adopted.
Eq. (\ref{qyb}) has a familiar graphic representation, which we omit 
for conciseness.

The consistency conditions implied by the scattering of two particles 
among each other
and with the impurity are conveniently organized in the following 
three groups:

(a) pure reflection:
\bea
&S_{12}(\chi_1, \chi_2)R^+_2(\chi_1)S_{12}(\chi_2 , 
-\chi_1)R^+_2(\chi_2) = \nonumber \\
&R^+_2(\chi_2)S_{12}(\chi_1, -\chi_2)R^+_2(\chi_1)S_{12}(-\chi_2, -\chi_1) \, ,
\label{SRSR+}
\eea
\bea
&S_{12}(\chi_1, \chi_2)R^-_1(\chi_2)S_{12}(-\chi_2 , 
\chi_1)R^-_1(\chi_1) = \nonumber \\
&R^-_1(\chi_1)S_{12}(-\chi_1, \chi_2)R^-_1(\chi_2)S_{12}(-\chi_2, -\chi_1) \, .
\label{SRSR-}
\eea
Eqs. (\ref{SRSR+}) and (\ref{SRSR-}) concern the reflection on $\hlp$ and
$\hlm$ respectively. Using the rules in Fig. 1 and moving back in time, one
gets the graphic representation of (\ref{SRSR+}) shown in Fig. 2.


\begin{figure}[h]
\setlength{\unitlength}{.5mm}
\begin{picture}(50,40)(-60,0)
\put(35,0){\line(0,1){40}}
\put(34,0){\line(0,1){40}}
\put(65,20){\line(-3,1){30}}
\put(50,25){\vector(-3,1){1}}
\put(35,30){\line(3,1){30}}
\put(50,35){\vector(3,1){1}}
\put(55,0){\line(-1,1){20}}
\put(45,10){\vector(-1,1){1}}
%
\put(35,20){\line(2,3){15}}
\put(47,38){\vector(2,3){1}}
\put(65,5){\makebox(20,15)[t]{=}}
%
\put(85,0){\line(0,1){40}}
\put(84,0){\line(0,1){40}}
%
\put(115,10){\line(-3,1){30}}
\put(109,12){\vector(-3,1){1}}
\put(85,20){\line(3,1){30}}
\put(100,25){\vector(3,1){1}}
%
\put(115,0){\line(-1,1){30}}
\put(105,10){\vector(-1,1){1}}
\put(85,30){\line(1,1){15}}
\put(95,40){\vector(1,1){1}}
\end{picture}
\caption{Pure reflection.}
\end{figure}
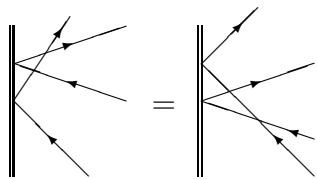

\noindent The picture associated to (\ref{SRSR-}) is obtained from 
Fig. 2 by reflection with respect
to the impurity world line.

(b) pure transmission:
\be
T^+_1(\chi_1)S_{12}(\chi_1, \chi_2)T^-_1(\chi_2) =
T^-_2(\chi_2)S_{12}(\chi_1, \chi_2)T^+_2(\chi_1) \, ,
\label{TST}
\ee
\be
S_{12}(\chi_1, \chi_2)T^-_1(\chi_2)T^-_2(\chi_1) =
T^-_1(\chi_1)T^-_2(\chi_2)S_{12}(\chi_1, \chi_2)\, ,
\label{STT-}
\ee
\be
S_{12}(\chi_1, \chi_2)T^+_1(\chi_2)T^+_2(\chi_1) =
T^+_1(\chi_1)T^+_2(\chi_2)S_{12}(\chi_1, \chi_2)\, ,
\label{STT+}
\ee
Eqs. (\ref{TST}) and (\ref{STT-}) are represented in Fig. 3(a) and 
Fig. 3(b) respectively.


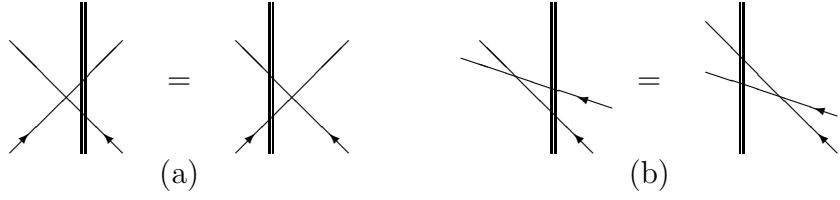
\begin{figure}[h]
\setlength{\unitlength}{.5mm}
\begin{picture}(40,40)(-25,-10)
\put(150,0){\line(0,1){40}}
\put(149,0){\line(0,1){40}}
\put(165,12){\line(-3,1){40}}
\put(156,15){\vector(-3,1){0.1}}
\put(160,0){\line(-1,1){30}}
\put(155,5){\vector(-1,1){0.1}}
\put(165,5){\makebox(20,15)[t]{=}}
\put(165,-10){\makebox(20,15)[b]{(b)}}
%
\put(200,0){\line(0,1){40}}
\put(199,0){\line(0,1){40}}
%
\put(225,10){\line(-3,1){35}}
\put(219,12){\vector(-3,1){0.1}}
%
\put(225,0){\line(-1,1){35}}
\put(220,5){\vector(-1,1){0.1}}
%
%
\put(25,0){\line(0,1){40}}
\put(24,0){\line(0,1){40}}
\put(5,0){\line(1,1){30}}
\put(10,5){\vector(1,1){0.1}}
\put(35,0){\line(-1,1){30}}
\put(30,5){\vector(-1,1){0.1}}
\put(40,5){\makebox(20,15)[t]{=}}
\put(40,-10){\makebox(20,15)[b]{(a)}}
%
\put(75,0){\line(0,1){40}}
\put(74,0){\line(0,1){40}}
%
\put(65,0){\line(1,1){30}}
\put(70,5){\vector(1,1){0.1}}
%
\put(95,0){\line(-1,1){30}}
\put(90,5){\vector(-1,1){0.1}}
\end{picture}
\caption{Pure transmission.}
\end{figure}

\noindent As before, the picture corresponding to eq. (\ref{STT+}) is 
obtained from Fig. 3(b) by reflection.

(c) mixed relations:
\be
R^+_1(\chi_1)T^-_2(\chi_2) =
T^-_2(\chi_2)S_{12}(\chi_1, \chi_2)R^+_2(\chi_1)S_{12}(\chi_2, -\chi_1) \, ,
\label{TSRS+}
\ee
\be
T^+_1(\chi_1)R^-_2(\chi_2) =
T^+_1(\chi_1)S_{12}(\chi_1, \chi_2)R^-_1(\chi_2)S_{12}(-\chi_2, \chi_1) \, ,
\label{TSRS-}
\ee
\be
R^+_1(\chi_1)T^+_2(\chi_2) =
S_{12}(\chi_1, \chi_2)R^+_2(\chi_1)S_{12}(\chi_2, -\chi_1)T^+_2(\chi_2) \, ,
\label{SRST+}
\ee
\be
T^-_1(\chi_1)R^-_2(\chi_2) =
S_{12}(\chi_1, \chi_2)R^-_1(\chi_2)S_{12}(-\chi_2, \chi_1)T^-_1(\chi_1)\, ,
\label{SRST-}
\ee
\be
R^+_1(\chi_1)T^-_2(\chi_2)S_{12}(-\chi_1, \chi_2) =
T^-_2(\chi_2)S_{12}(\chi_1, \chi_2)R^+_2(\chi_1) \, ,
\label{TSR+}
\ee
\be
T^+_1(\chi_1)R^-_2(\chi_2)S_{12}(\chi_1, -\chi_2) =
T^+_1(\chi_1)S_{12}(\chi_1, \chi_2)R^-_1(\chi_2) \, ,
\label{TSR-}
\ee
\be
R^+_2(\chi_1)S_{12}(\chi_2, -\chi_1)T^+_2(\chi_2) =
S_{12}(\chi_2, \chi_1)R^+_1(\chi_1)T^+_2(\chi_2) \, ,
\label{RST+}
\ee
\be
R^-_1(\chi_2)S_{12}(-\chi_2, \chi_1)T^-_1(\chi_1) =
S_{12}(\chi_2, \chi_1)T^-_1(\chi_1)R^-_2(\chi_2) \, .
\label{RST-}
\ee
Eqs. (\ref{TSRS+}) and (\ref{TSR+}) are shown in Fig. 4(a) and 4(b) 
respectively,
whereas eqs. (\ref{SRST+}) and (\ref{RST+}) are drown in Fig. 4(c) and 4(d).
The pictures related to the remaining four mixed equations are 
obtained from Fig. 4 by
reflection. This completes our discussion of the three-body processes.

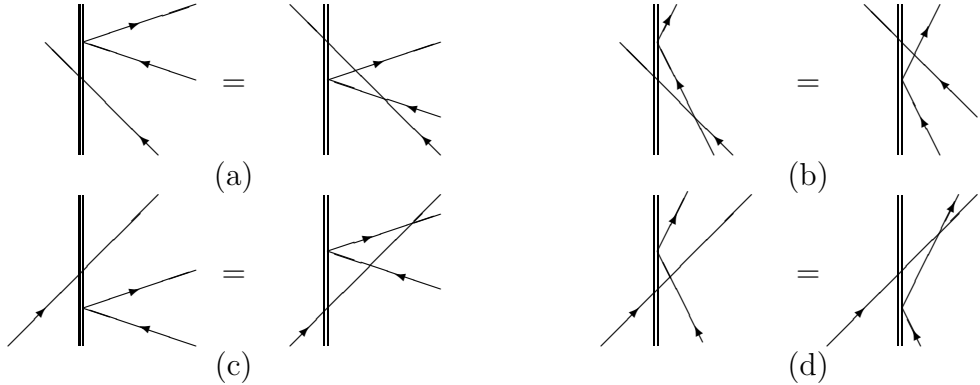
\begin{figure}[ht]
\setlength{\unitlength}{.5mm}
\begin{picture}(120,50)(10,-10)
\put(35,0){\line(0,1){40}}
\put(34,0){\line(0,1){40}}
\put(65,20){\line(-3,1){30}}
\put(50,25){\vector(-3,1){0.1}}
\put(35,30){\line(3,1){30}}
\put(50,35){\vector(3,1){0.1}}
\put(55,0){\line(-1,1){30}}
\put(50,5){\vector(-1,1){0.1}}
\put(65,5){\makebox(20,15)[t]{=}}
\put(65,-10){\makebox(20,15)[b]{(a)}}
%
\put(100,0){\line(0,1){40}}
\put(99,0){\line(0,1){40}}
%
\put(130,10){\line(-3,1){30}}
\put(121,13){\vector(-3,1){0.1}}
\put(100,20){\line(3,1){30}}
\put(115,25){\vector(3,1){0.1}}
%
\put(130,0){\line(-1,1){40}}
\put(125,5){\vector(-1,1){0.1}}
\end{picture}
%
%
\begin{picture}(120,50)(-20,-10)
\put(35,0){\line(0,1){40}}
\put(34,0){\line(0,1){40}}
\put(50,0){\line(-1,2){15}}
\put(40,20){\vector(-1,2){0.1}}
\put(35,30){\line(1,2){5}}
\put(38.5,37){\vector(1,2){0.1}}
\put(55,0){\line(-1,1){30}}
\put(50,5){\vector(-1,1){0.1}}
\put(65,5){\makebox(20,15)[t]{=}}
\put(65,-10){\makebox(20,15)[b]{(b)}}
%
\put(100,0){\line(0,1){40}}
\put(99,0){\line(0,1){40}}
%
\put(110,0){\line(-1,2){10}}
\put(105,10){\vector(-1,2){0.1}}
\put(100,20){\line(1,2){10}}
\put(107,34){\vector(1,2){0.1}}
%
\put(120,10){\line(-1,1){30}}
\put(110,20){\vector(-1,1){0.1}}
\end{picture}


\begin{picture}(140,50)(10,-10)
\put(35,0){\line(0,1){40}}
\put(34,0){\line(0,1){40}}
%
\put(15,0){\line(1,1){40}}
\put(25,10){\vector(1,1){0.1}}
\put(65,0){\line(-3,1){30}}
\put(50,5){\vector(-3,1){0.1}}
\put(35,10){\line(3,1){30}}
\put(50,15){\vector(3,1){0.1}}
\put(65,5){\makebox(20,15)[t]{=}}
\put(65,-10){\makebox(20,15)[b]{(c)}}
%
\put(100,0){\line(0,1){40}}
\put(99,0){\line(0,1){40}}
%
\put(130,15){\line(-3,1){30}}
\put(118,19){\vector(-3,1){0.1}}
\put(100,25){\line(3,1){30}}
\put(112,29){\vector(3,1){0.1}}
\put(90,0){\line(1,1){40}}
\put(95,5){\vector(1,1){0.1}}
\end{picture}
%
%
\begin{picture}(120,50)(0,-10)
\put(35,0){\line(0,1){40}}
\put(34,0){\line(0,1){40}}
\put(47,1){\line(-1,2){12}}
\put(45,5){\vector(-1,2){0.1}}
\put(35,25){\line(1,2){8}}
\put(40,35){\vector(1,2){0.1}}
\put(20,0){\line(1,1){40}}
\put(30,10){\vector(1,1){0.1}}
\put(65,5){\makebox(20,15)[t]{=}}
\put(65,-10){\makebox(20,15)[b]{(d)}}
%
\put(100,0){\line(0,1){40}}
\put(99,0){\line(0,1){40}}
%
\put(80,0){\line(1,1){40}}
\put(90,10){\vector(1,1){0.1}}
\put(105,0){\line(-1,2){5}}
\put(103,4){\vector(-1,2){0.1}}
\put(100,10){\line(1,2){15}}
\put(114,38){\vector(1,2){0.1}}
\end{picture}

\caption{Mixed relations.}
\end{figure}

Let us focus finally on the requirements of unitarity and Hermitian 
analyticity. For $S$
one has the familiar \cite{Cherednik:vs, Sklyanin:yz, Ghoshal:tm} conditions
\be
S_{12}(\chi_1,\chi_2) S_{12}(\chi_2,\chi_1) = 1 \, ,
\label{uS}
\ee
\be
[S_{12}]^\dagger (\chi_1,\chi_2) = S_{12}(\chi_2,\chi_1) \, ,
\label{haS}
\ee
where the dagger stands for Hermitian conjugation. Concerning $R$ and $T$,
the conditions of unitarity and Hermitian analyticity read
\be
T(\chi) T(\chi) + R (\chi) R (-\chi) = 1 \, ,
\label{uTTRR}
\ee
\be
T (\chi) R (\chi) + R (\chi) T (-\chi) = 0 \, ,
\label{uTRTR}
\ee
and
\be
{[T]}^\dagger (\chi) = T(\chi)\, ,\qquad {[R]}^\dagger (\chi) = R (-\chi)\, ,
\label{haTR}
\ee
respectively. We will come back to eqs. (\ref{uTTRR}-\ref{haTR}) in
Sect. 5, observing for the moment that (\ref{uS}) implies the 
equivalence of the two sets
of equations (\ref{TSRS+}-\ref{SRST-}) and (\ref{TSR+}-\ref{RST-}). Therefore
one is left with the study eqs. (\ref{qyb}-\ref{SRST-}) and 
(\ref{uS}-\ref{haTR}),
which is the main subject of the next section.

\sect{Solutions of the consistency relations}

Let us assume in what follows that $S$ obeys (\ref{qyb}, \ref{uS}, \ref{haS}).
Our aim below is to find the general solution of
eqs. (\ref{SRSR+}-\ref{RST-}) when the
conditions of unitarity and Hermitian analyticity (\ref{uTTRR}-\ref{haTR})
are satisfied and the matrix $T$
is invertible.  We shall present also some families of non trivial solutions,
which are of physical interest.

Because of (\ref{haTR}),
$T(\chi) T(\chi)$ and $R (\chi) R (-\chi)$ are non negative Hermitian 
matrices. From
(\ref{uTTRR}) it follows that they are diagonalizable simultaneously 
and that the corresponding
eigenvalues satisfy
\be
\lambda_i(\chi) + \mu_i(\chi) = 1\, ,\qquad \lambda_i(\chi) \geq 0\, ,
\quad \mu_i(\chi) \geq 0\, , \qquad i=1,...,N.
\label{eig}
\ee
Now, using eq. (\ref{uTTRR}), one can express $T$ as a function of
$R$:
\be
T(\chi)=t(\chi)\sqrt{1-R(\chi)[R(\chi)]^\dag}=t(\chi)\sum_{n=0}^\infty
\alpha_{n}\,\Big(R(\chi)R(-\chi)\Big)^n\, ,
\label{T=eps(R)}
\ee
where $t(\chi)\in\{-1,+1\}$ is some unknown function and the real
numbers $\alpha_{n}$ are defined through the expansion
$\sqrt{1-x}=\sum_{n=0}^\infty \alpha_{n}x^n$. Notice, that the
conditions (\ref{eig}) ensure that this series is convergent.
Demanding $T$ to satisfy eq. (\ref{uTRTR}) shows that
$t(\chi)$ must be an odd function. If
$t(\chi)$ is in addition continuous for $\chi\not=0$, one concludes that
$t(\chi)=\pm \eps(\chi)$, where $\eps$ is the sign function.

In the following we assume that $T$ is invertible, which is equivalent to
$\lambda_{i}(\chi)\neq0$ for all $i=1,...N$. Physically, this means 
that there is no
isotopic channel with pure reflection only. If such channels exist, 
one can separate
the corresponding isotopic degrees of freedom and treat them as a system with
pure reflection. For invertible $T$, one gets from eqs. (\ref{TSRS+}) and
(\ref{TSRS-}) that the matrices $R^\pm$ obey
$S_{12}(\chi_1,\chi_2)R^\pm_2(\chi_1) = R^\pm_1(\chi_1)S_{12}(-\chi_1,\chi_2)$,
which can be compactly rewritten as
\be
S_{12}(\chi_1,\chi_2)R_2(\chi_1) = R_1(\chi_1)S_{12}(-\chi_1,\chi_2) \, .
\label{SR}
\ee
It is a simple matter to prove that (\ref{SR}) solves all the
mixed relations (\ref{TSRS+}-\ref{RST-}), as well as the pure reflection ones
(\ref{SRSR+},\ref{SRSR-}). Moreover, from
the expression (\ref{T=eps(R)}) and the eq. (\ref{SR}), one easily
deduces that
\be
S_{12}(\chi_1,\chi_2)T_2(\chi_1) = T_1(\chi_1)S_{12}(\chi_1,\chi_2) \, .
\label{ST}
\ee
Finally, a direct inspection shows that any matrix $T$ satisfying eq.
(\ref{ST}) is a solution to  the pure transmission eqs.
(\ref{TST}-\ref{STT+}).

Let us stress that, when the unitarity
conditions (\ref{uS}-\ref{uTRTR}) and the invertibility of $T$ are
assumed, the whole set of equations (\ref{SRSR+}-\ref{RST-}) is
equivalent to the two simple linear equations (\ref{SR}) and (\ref{ST}).
Summarizing, we proved that the Hermitian matrix
\be
T(\chi)=\pm \eps(\chi)\sqrt{1-R(\chi)[R(\chi)]^\dag}
\label{T=sgn(R)}
\ee
with $R$ obeying eq. (\ref{SR}), is the general solution to the eqs. 
(\ref{SRSR+}-\ref{RST-}).

Collecting the results of this section, we have reduced the original problem
to the solution of
eqs.(\ref{uTTRR}-\ref{haTR}) and (\ref{SR},\ref{ST}).
It is instructive at this point to produce some explicit examples.
We start with the $gl(N)$-invariant $S$-matrix
\be
S_{12}(\chi_1,\chi_2)=\frac{1}{s(\chi_1) - s(\chi_2) +i\, g}
\left\{\left [s(\chi_1) - s(\chi_2)\right ]\, P_{{12}}+ig \, 
\II\otimes \II \right\} \, ,
\label{Sgln}
\ee
where $P_{12}$ is the standard flip operator,
$\II$ is the $N\times N$ identity matrix,
$g\in \RR$ and $s(\chi)$ is any real valued
{\sl even} function. For $R$ and $T$ one easily derives
\be
R(\chi ) = [\cos p(\chi )]\exp [iq(\chi)]\, \II\, ,\qquad
T(\chi ) = [\sin p(\chi )]\, \II \, ,
\label{RTgln}
\ee
$p(\chi )$ and $q(\chi )$ being real valued {\sl odd} functions. In 
this example
both reflection and transmission preserve the isotopic type. Moreover,
all isotopic types have the same reflection and transmission coefficient.

Slightly more complicated is the Toda type S-matrix
\be
S_{i_1i_2}^{j_1j_2}(\chi_1,\chi_2) = \exp\left 
[is_{i_1i_2}(\chi_1,\chi_2)\right ]\,
\delta_{i_1}^{j_2} \delta_{i_2}^{j_1} \, ,
\label{Stoda}
\ee
where $s_{i_1i_2}(\chi_1,\chi_2)$ are real valued functions obeying
\be
s_{i_1i_2}(\chi_1,\chi_2) = - s_{i_2i_1}(\chi_2,\chi_1)\, , \qquad
s_{i_1i_2}(\chi_1,\chi_2) = s_{i_1i_2}(\chi_1,-\chi_2)\, .
\label{todas}
\ee
If we assume furthermore that for any couple of indices $(i,j)$ with $i\neq j$,
there exists an index $k$ such that $s_{ik}(\chi_1,\chi_2) $ is
different from $s_{jk}(\chi_1,\chi_2)$,
one finds
\be
R_i^j(\chi ) = [\cos p_i(\chi )]\exp [iq_i(\chi)]\, \delta_i^j\, , \qquad
T_i^j(\chi ) = [\sin p_i(\chi )]\, \delta_i^j\, ,
\label{todaRT}
\ee
with all $p_i(\chi )$ and $q_i(\chi )$ real valued {\sl odd} functions.
Also here the impurity interaction preserves the isotopic type, but 
the individual
reflection and transmission coefficients may be different.
 
On the other hand, whenever a couple of indices $(i_{0},j_{0})$
with $i_{0}\neq j_{0}$ exists,
such that $s_{i_{0}k}(\chi_1,\chi_2) = s_{j_{0}k}(\chi_1,\chi_2)$
for any $k$, one has in general that
the corresponding off-diagonal matrix elements $R_{i_{0}}^{j_{0}}(\chi)$ and
$R_{j_{0}}^{i_{0}}(\chi)$, as well as $T_{i_{0}}^{j_{0}}(\chi)$ and
$T_{j_{0}}^{i_{0}}(\chi)$, do not vanish. They describe
impurity interactions, which does not preserve the isotopic type.

\sect{Boundary algebra with transmission\label{sectboundalg}}

The goal of this section is to provide an algebraic framework for
integrable models with impurity, analogous to the 
Zamolodchikov-Faddeev (ZF) algebra
$\A$ operating in the case without impurities.
The main idea is to adapt
the concept of boundary algebra $\B$ \cite{Liguori:1998xr} to the case
in which the boundary is both reflecting and transmitting.
It might be useful to recall in this respect that in the case
with pure reflection all scattering processes take place only in
$\RR_+$ (or $\RR_-$).
The whole line $\RR$ is involved instead, if nontrivial transmission 
is present.
Moreover, the impurity
at $x=0$ breaks down translation invariance. In order to take into
account these two facts, we equip the generators of our algebra
with a double index $\alpha = (\xi , i)$. As before,
the index $i=1,...,N$ denotes the isotopic type. The index $\xi = \pm$
indicates the half line $\RR_\pm$ where the particle is created or annihilated.
With this notation, we consider an associative algebra $\C$ with 
identity element $\bf 1$,
generated by $\{a_\alpha (\chi ),\, a^{\ast \alpha } (\chi )\}$, 
which are subject to
the constraints:
\bea
a_{\alpha_1}(\chi_1)\, a_{\alpha_2 }(\chi_2) \, \; - \; \,
      {\cal S}_{\alpha_2 \alpha_1 }^{\beta_1 \beta_2 }
      (\chi_2 , \chi_1)\, a_{\beta_2 }(\chi_2)\, a_{\beta_1 }(\chi_1) & = & 0
      \, , \qquad \label{aa}\\
a^{\ast \alpha_1 } (\chi_1)\, a^{\ast \alpha_2 } (\chi_2) -
      a^{\ast \beta_2 } (\chi_2)\, a^{\ast \beta_1 } (\chi_1)\,
      {\cal S}_{\beta_2 \beta_1 }^{\alpha_1 \alpha_2 }(\chi_2 , \chi_1) & = & 0
      \, , \qquad \label{a*a*} \\
a_{\alpha_1 }(\chi_1)\, a^{\ast \alpha_2 } (\chi_2) \; - \;
      a^{\ast \beta_2 }(\chi_2)\,
      {\cal S}_{\alpha_1 \beta_2 }^{\alpha_2 \beta_1 }(\chi_1 , \chi_2)\,
      a_{\beta_1 }(\chi_1) & = &  \nonumber\\
      \delta (\chi_1 - \chi_2)
      \left [\delta_{\alpha_1 }^{\alpha_2 } + 
\T_{\alpha_1}^{\alpha_2}(\chi_1)\right]\, {\bf 1} +
      \delta (\chi_1 + \chi_2)\,  \R_{\alpha_1 }^{\alpha_2 }(\chi_1)\, {\bf 1}
      \, .
\label{aa*}
\eea
Here $\{{\cal S}, \R, \T\}$ are related to the starting data 
$\{S,R,T\}$ in the following way:
\be
{\cal S}_{\, (\xi_1,i_1)\, (\xi_2,i_2)}^{(\eta_1,j_1) (\eta_2,j_2)}
(\chi_1 , \chi_2) \equiv \delta_{\xi_1}^{\eta_2} \delta_{\xi_2}^{\eta_1} \,
S_{i_1 i_2}^{j_1 j_2} (\chi_1 , \chi_2) \, ,
\label{calS}
\ee
\be
\R_{(\xi,i)}^{(\eta,j)} (\chi ) \equiv \delta_\xi^\eta \, R_i^j (\chi 
) \, , \qquad
\T_{\, \, (\xi,i)}^{(\eta,j)} (\chi ) \equiv \varepsilon_\xi^\eta \, 
T_i^j (\chi ) \, ,
\label{calRT}
\ee
where
\be
\varepsilon =\left(\begin{array}{cc} 0&1\\ 1&0\end{array}\right)\, .
\label{epsilon}
\ee

This algebra differs from the one proposed in \cite{Delfino:1994nr, 
Konik:1997gx, Castro-Alvaredo:2002fc}.
Comparing $\C$ to the ZF algebra $\A$, the presence of the new terms 
$\R$ and $\T$ in
(\ref{aa*}) must be emphasized. Only the $\R$-term appears in the 
boundary algebra $\B$,
but in general as a new generator not proportional to the identity 
${\bf 1}$. This difference
of $\B$ with respect to $\C$ can be traced back to condition 
(\ref{SR}), which is stronger
than (\ref{SRSR+},\ref{SRSR-}). Let us remark also that the $\R$-term 
in (\ref{aa*}) breaks down
both Lorentz (Galilean) and translation invariance.

It is easy to verify that each triplet $\{S,R,T\}$, obeying
eqs. (\ref{qyb},\ref{uS}-\ref{haTR}) and (\ref{SR},\ref{ST}), determines
a triplet $\{{\cal S}, \R, \T\}$, which satisfies the same equations.
Using this fact and following
closely the formalism developed in \cite{Liguori:1998xr, Liguori:de},
one can construct the Fock representation $\F$ of $\C$.
Referring for the details to \cite{MRS},
let us collect here those basic features of $\F$, needed in the next 
section for
the definition of the scattering operator. After smearing,
$\{a_\alpha (\chi ),\, a^{\ast \alpha } (\chi )\}$ are
represented by densely defined
operators, acting in a Hilbert space $\Hil$ with scalar product 
$(\cdot\, ,\, \cdot)$.
There exist a cyclic vacuum state $\Omega \in \Hil$, which is annihilated by
$\{a_\alpha (\chi )\}$ and satisfies $(\Omega\, ,\, \Omega) = 1$.

The asymptotic states are prepared in $\Hil$ as follows. In-states 
are created from the vacuum
by $\{a^{\ast (-,i)}(\varphi )\, :\, \varphi >0\}$ and
$\{a^{\ast (+,i)}(\varphi )\, :\, \varphi < 0\}$.
The out-states are generated instead by
$\{a^{\ast (-,j)}(\chi )\, :\, \chi <0\}$ and
$\{a^{\ast (+,j)}(\chi )\, :\, \chi > 0\}$.
This choice corresponds to incoming
particles traveling towards the impurity and outgoing
particles moving in the opposite directions.
Without loss of generality \cite{Cherednik:vs, Sklyanin:yz}, one can 
also order the
creation operators using the values of the spectral parameter 
(rapidity). We thus
define
\be
|\varphi_1,\alpha_1;...;\varphi_m,\alpha_m\rangle^{\rm in} =
a^{\ast \alpha_1}(\varphi_1)\cdots a^{\ast \alpha_m}(\varphi_m ) \Omega \, ,
\label{in}
\ee
with
\be
\varphi_1<...<\varphi_m\, , \qquad \alpha_k = (\xi_k,i_k)\, ,\qquad
\xi_k=-\eps(\varphi_k)\, ,
\label{order-in}
\ee
$\eps$ being the sign function. Analogously
\be
{}^{\rm out}\langle \chi_1,\beta_1;...;\chi_n,\beta_n| =
a^{\ast \beta_1}(\chi_1)\cdots a^{\ast \beta_n}(\chi_n ) \Omega \, ,
\label{out}
\ee
where
\be
\chi_1>...>\chi_n\, , \qquad \beta_l = (\eta_l, j_l)\, ,\qquad \eta_l 
=  \eps(\chi_l)\, .
\label{order-out}
\ee
The asymptotic spaces $\F^{\rm in}$ and $\F^{\rm out}$ are generated by finite
linear combinations of vectors of the type (\ref{in}) and (\ref{out}) 
respectively.
Each of these spaces is dense in $\Hil$, ensuring asymptotic completeness.

Finally, one can express via
\bea
{}^{\rm out}\langle \chi_1,\beta_1;...;\chi_n,\beta_n
|\varphi_1,\alpha_1;...;\varphi_m,\alpha_m\rangle^{\rm in} = \\ \nonumber
\left (a^{\ast \beta_1}(\chi_1)\cdots a^{\ast \beta_n}(\chi_n ) \Omega \, ,
a^{\ast \alpha_1}(\varphi_1)\cdots a^{\ast \alpha_m}(\varphi_m ) 
\Omega \right )
\label{trampl}
\eea
a generic scattering amplitude in terms of the correlation functions,
which can be computed in turn by means of the exchange relation (\ref{aa*}).
The Fock structure implies that this amplitude vanishes unless $m=n$,
which corresponds physically to the absence of particle production 
due to integrability.

\sect{The total scattering operator}

In order to become more familiar with the scattering
theory, encoded in the representation $\F$ of $\C$,
we derive now some transition amplitudes in explicit form. The 
simplest ones are
the one-particle amplitudes. All of them can be deduced from
the correlation function
\bea
&\left (a^{\ast \beta}(\chi)\Omega \, , \,
a^{\ast \alpha}(\vph) \Omega \right ) =   \nonumber \\
&\left [\delta^{\alpha}_{\beta}+
{{\T}}^{\alpha}_{\beta}(\chi)\right ]\delta(\chi-\vph)+
{{\R}}^{\alpha}_{\beta}(\chi) \delta(\chi+\varphi) \, .
\label{two-point function}
\eea
Taking into account eqs. (\ref{order-in},\ref{order-out}),
one has the following four possibilities:
\be
{}^{\rm out}\langle \chi,(\eta,j)|\vph,(\xi,i)\rangle^{\rm in}  =
\left\{ \begin{array}{cc}
{T^+}^{i}_{j}(\chi)\delta(\chi-\varphi), & \quad \mbox{$\xi=+,\, \eta=+$}\, ,
\\ {R^-}^{i}_{j}(\chi)\delta(\chi+\varphi), & \quad \mbox{$\xi=+,\, 
\eta=-$}\, ,
\\ {R^+}^{i}_{j}(\chi)\delta(\chi+\varphi), & \quad \mbox{$\xi=-,\, 
\eta=+$}\, ,
\\ {T^-}^{i}_{j}(\chi)\delta(\chi-\varphi), & \quad \mbox{$\xi=-,\, 
\eta=-$}\, .
\end{array} \right.
\label{one-particle}
\end{equation}
These amplitudes have transparent physical interpretation and describe the
particle-impurity interaction. Notice that eqs. (\ref{uTTRR}-\ref{haTR}) ensure
one-particle unitarity.

The particle-particle interaction shows up
in the two-particle amplitudes, which can be deduced from the correlator
\bea
\lefteqn{
\left (a^{\ast \beta_1}(\chi_1) a^{\ast \beta_2}(\chi_2 ) \Omega \, ,
\, a^{\ast \alpha_1}(\vph_1) a^{\ast \alpha_2}(\vph_2 ) \Omega \right 
) \ =\   }\nonu
&&[\delta^\mu_{\beta_{2}}+{{\T}}^\mu_{\beta_{2}}(\chi_{2})]\,
{\cS}_{\beta_{1}\mu}^{\alpha_{1}\nu}(\chi_{1},\chi_{2})\,
[\delta^{\alpha_{2}}_{\nu}+{{\T}}^{\alpha_{2}}_{\nu}(\chi_{1})]\
\delta(\chi_{1}-\varphi_{2})\, \delta(\chi_{2}-\vph_{1})\nonu
&+&{{\R}}^\mu_{\beta_{2}}(\chi_{2})\,
{\cS}_{\beta_{1}\mu}^{\alpha_{1}\nu}(\chi_{1},-\chi_{2})\,
[\delta^{\alpha_{2}}_{\nu}+{{\T}}^{\alpha_{2}}_{\nu}(\chi_{1})]\
\delta(\chi_{1}-\varphi_{2})\, \delta(\chi_{2}+\vph_{1})\nonu
&+&[\delta^\mu_{\beta_{2}}+{{\T}}^\mu_{\beta_{2}}(\chi_{2})]\,
{{\cS}}_{\beta_{1}\mu}^{\alpha_{1}\nu}(\chi_{1},\chi_{2})\,
{{\R}}^{\alpha_{2}}_{\nu}(\chi_{1})\
\delta(\chi_{1}+\varphi_{2})\, \delta(\chi_{2}-\vph_{1})\nonu
&+&{{\R}}^\mu_{\beta_{2}}(\chi_{2})\,
{{\cS}}_{\beta_{1}\mu}^{\alpha_{1}\nu}(\chi_{1},-\chi_{2})\,
{{\R}}^{\alpha_{2}}_{\nu}(\chi_{1})\
\delta(\chi_{1}+\varphi_{2})\, \delta(\chi_{2}+\vph_{1})\nonu
&+&[\delta^{\alpha_1}_{\beta_1}+\T^{\alpha_1}_{\beta_1}(\chi_1)]\,
[\delta^{\alpha_2}_{\beta_2}+\T^{\alpha_2}_{\beta_2}(\chi_2)]\
\delta(\chi_1-\varphi_1)\, \delta(\chi_2-\vph_2)\nonu
&+&[\delta^{\alpha_1}_{\beta_1}+\T^{\alpha_1}_{\beta_1}(\chi_1)]\,
\R^{\alpha_2}_{\beta_2}(\chi_2)\
\delta(\chi_1-\varphi_1)\, \delta(\chi_2+\vph_2)\nonu
&+&\R^{\alpha_1}_{\beta_1}(\chi_1)\,
[\delta^{\alpha_2}_{\beta_2}+\T^{\alpha_2}_{\beta_2}(\chi_2)]\
\delta(\chi_1+\varphi_1)\, \delta(\chi_2-\vph_2)\nonu
&+&\R^{\alpha_1}_{\beta_1}(\chi_1)\,
\R^{\alpha_2}_{\beta_2}(\chi_2)\
\delta(\chi_1+\varphi_1)\, \delta(\chi_2+\vph_2)\, .
\label{4-point}
\eea
Using the definition (\ref{in}-\ref{order-out}) of asymptotic states, one has
various kinematic domains, depending on the sign of $\varphi_i$ and $\chi_j$.
There are four cases with
$\eps (\varphi_1)=\eps (\varphi_2)$ and $\eps (\chi_1)=\eps (\chi_2)$,
the corresponding amplitudes being:
\bea
{}^{\rm out}\langle 
\chi_1,(\pm,j_1);\chi_2,(\pm,j_2)|\vph_1,(\pm,i_1);\vph_2,(\pm,i_2)
\rangle^{\rm in} = \nonumber \\
{{T^\pm}}^k_{j_{2}}(\chi_{2})\, {S}_{j_{1} k}^{i_{1}l}(\chi_{1},\chi_{2})\,
{{T^\pm}}^{i_{2}}_{l}(\chi_{1})\
\delta(\chi_{1}-\varphi_{2})\, \delta(\chi_{2}-\vph_{1})\, ,
\label{a1}
\eea
\bea
{}^{\rm out}\langle 
\chi_1,(\pm,j_1);\chi_2,(\pm,j_2)|\vph_1,(\mp,i_1);\vph_2,(\mp,i_2)
\rangle^{\rm in} = \nonumber \\
{R^\pm}^{i_1}_{j_1}(\chi_1)\,
{R^\pm}^{i_2}_{j_2}(\chi_2)\
\delta(\chi_1+\varphi_1)\, \delta(\chi_2+\vph_2)\, ,
\eea
There exist other four cases in which
$\eps (\varphi_1)\not=\eps (\varphi_2)$ and $\eps (\chi_1)=\eps (\chi_2)$ or
$\eps (\varphi_1)=\eps (\varphi_2)$ and $\eps (\chi_1)\not=\eps (\chi_2)$. From
eq. (\ref{4-point}) one gets the amplitudes:
\bea
{}^{\rm out}\langle \chi_1,(+,j_1);\chi_2,(+,j_2)|\vph_1,(-,i_1);\vph_2,(+,i_2)
\rangle^{\rm in} = \nonumber \\
{{R^+}}^k_{j_{2}}(\chi_{2}){S}_{j_{1}k}^{i_{1}l}(\chi_{1},-\chi_{2})
{{T^+}}^{i_{2}}_{l}(\chi_{1})\delta(\chi_{1}-\varphi_{2})\delta(\chi_{ 
2}+\vph_{1})
\nonumber \\ +
{R^+}^{i_1}_{j_1}(\chi_1){T^+}^{i_2}_{j_2}(\chi_2)\
\delta(\chi_1+\varphi_1)\delta(\chi_2-\vph_2) \, ,
\eea
\bea
{}^{\rm out}\langle \chi_1,(-,j_1);\chi_2,(-,j_2)|\vph_1,(-,i_1);\vph_2,(+,i_2)
\rangle^{\rm in} = \nonumber \\
{{T^-}}^k_{j_{2}}(\chi_{2}){S}_{j_{1}k}^{i_{1}l}(\chi_{1},\chi_{2})
{{R^-}}^{i_{2}}_{l}(\chi_{1})\delta(\chi_{1}+\varphi_{2})\delta(\chi_{ 
2}-\vph_{1})
\nonumber \\ +
{T^-}^{i_1}_{j_1}(\chi_1){R^-}^{i_2}_{j_2}(\chi_2)\
\delta(\chi_1-\varphi_1)\delta(\chi_2+\vph_2) \, ,
\eea
\bea
{}^{\rm out}\langle \chi_1,(+,j_1);\chi_2,(-,j_2)|\vph_1,(+,i_1);\vph_2,(+,i_2)
\rangle^{\rm in} = \nonumber \\
{{R^-}}^k_{j_{2}}(\chi_{2}){S}_{j_{1}k}^{i_{1}l}(\chi_{1},-\chi_{2})
{{T^+}}^{i_{2}}_{l}(\chi_{1})\delta(\chi_{1}-\varphi_{2})\delta(\chi_{ 
2}+\vph_{1})
\nonumber \\ +
{T^+}^{i_1}_{j_1}(\chi_1){R^-}^{i_2}_{j_2}(\chi_2)\
\delta(\chi_1-\varphi_1)\delta(\chi_2+\vph_2) \, ,
\eea
\bea
{}^{\rm out}\langle \chi_1,(+,j_1);\chi_2,(-,j_2)|\vph_1,(-,i_1);\vph_2,(-,i_2)
\rangle^{\rm in} = \nonumber \\
{{T^-}}^k_{j_{2}}(\chi_{2}){S}_{j_{1}k}^{i_{1}l}(\chi_{1},\chi_{2})
{{R^+}}^{i_{2}}_{l}(\chi_{1})\delta(\chi_{1}+\varphi_{2})\delta(\chi_{ 
2}-\vph_{1})
\nonumber \\ +
{R^+}^{i_1}_{j_1}(\chi_1){T^-}^{i_2}_{j_2}(\chi_2)\
\delta(\chi_1+\varphi_1)\delta(\chi_2-\vph_2) \, .
\eea
Finally, the case $\eps (\varphi_1)\not=\eps (\varphi_2)$ and
$\eps (\chi_1)\not=\eps(\chi_2)$ exhausts all possibilities. The 
corresponding amplitude is
\bea
{}^{\rm out}\langle \chi_1,(+,j_1);\chi_2,(-,j_2)|\vph_1,(-,i_1);\vph_2,(+,j_2)
\rangle^{\rm in} = \nonumber \\
{{T^-}}^k_{j_{2}}(\chi_{2}){S}_{j_{1}k}^{i_{1}l}(\chi_{1},\chi_{2})
{{T^+}}^{i_{2}}_{l}(\chi_{1})\delta(\chi_{1}-\varphi_{2})\delta(\chi_{ 
2}-\vph_{1})
\nonumber \\ +
{R^+}^{i_1}_{j_1}(\chi_1){R^-}^{i_2}_{j_2}(\chi_2)\
\delta(\chi_1+\varphi_1)\delta(\chi_2+\vph_2) \, .
\label{a9}
\eea
Keeping in mind that $R$ and $T$ satisfy eqs. (\ref{SR},\ref{ST}), 
one can recover
from eqs. (\ref{a1}-\ref{a9}) the analytic expressions of all 
two-particle processes,
some of which represented by the diagrams in Figs. 2-4.

Any $n$-particle scattering amplitude can be reconstructed with the 
above algorithm.
These amplitudes (\ref{trampl}) define the total
scattering operator\break ${\bf S}\, :\, \F^{\rm out}\to \F^{\rm 
in}$, which acts according to
\be
{\bf S}\, :\, a^{\ast \beta_1}(\chi_1)\cdots a^{\ast \beta_n}(\chi_n ) \Omega
\longmapsto a^{\ast {\widetilde\beta}_n}(\chi_n)\cdots a^{\ast 
{\widetilde\beta}_1}(\chi_1 ) \Omega \, ,
\ee
where $\chi_1>...>\chi_n$ and $\beta_l = (\eta_l, j_l),\, 
{\widetilde\beta}_l = (-\eta_l, j_l)$.
One can check that
\be
\left({\bf S}\Psi^{\rm out}\, ,\, {\bf S}\Phi^{\rm out}\right) =
\left(\Psi^{\rm out}\, ,\, \Phi^{\rm out}\right) \,
\ee
holds on $\F^{\rm out}$, which together with asymptotic completeness 
and invertibility of $\bf S$, implies its
unitarity.

\sect{Outlook and conclusions}

Factorized scattering theory in the presence of a reflecting and 
transmitting impurity has been
investigated. We have shown that relaxing the condition on the bulk 
scattering matrix
to depend only on the difference of the spectral parameters of the 
colliding particles, allows
for non trivial solutions of the three-body consistency relations. We 
established the general one
in the case of invertible transmission factor. Our philosophy in
constructing the scattering amplitudes is not to postulate the existence of a
boundary state with certain reflection and transmission properties. 
We rather prefer
to deal with an algebra, which admits a Fock representation, whose 
cyclic (vacuum) state
plays the role of boundary state. This approach has already shown 
some advantages
in the derivation of off-shell correlation functions 
\cite{Gattobigio:1998hn, Gattobigio:1998si} and
the study of symmetries \cite{Mintchev:2001aq}. We demonstrate above 
that it works
also in the case of impurities and provides a direct and relatively 
simple construction
of the total scattering operator. A further generalization of the
algebraic structure, applied in this paper, is presently under 
investigation \cite{MRS}.


\begin{thebibliography}{99}

\bibitem{Cherednik:vs}
I.~V.~Cherednik,
Theor.\ Math.\ Phys.\  {\bf 61} (1984) 977
[Teor.\ Mat.\ Fiz.\  {\bf 61} (1984) 35].

\bibitem{Sklyanin:yz}
E.~K.~Sklyanin,
J.\ Phys.\ A {\bf 21} (1988) 2375.

\bibitem{Ghoshal:tm}
S.~Ghoshal and A.~B.~Zamolodchikov,
Quantum Field Theory,''
Int.\ J.\ Mod.\ Phys.\ A {\bf 9} (1994) 3841
[Erratum-ibid.\ A {\bf 9} (1994) 4353]
[arXiv:hep-th/9306002].

\bibitem{Saleur:1998hq}
H.~Saleur,
``Lectures on non perturbative field theory and quantum impurity  problems,''
arXiv:cond-mat/9812110.

\bibitem{Saleur:2000gp}
H.~Saleur,
``Lectures on non perturbative field theory and quantum impurity 
problems. II,''
arXiv:cond-mat/0007309.

\bibitem{Delfino:1994nr}
G.~Delfino, G.~Mussardo and P.~Simonetti,
with a line of defect,''
Nucl.\ Phys.\ B {\bf 432} (1994) 518
[arXiv:hep-th/9409076].

\bibitem{Konik:1997gx}
R.~Konik and A.~LeClair,
Nucl.\ Phys.\ B {\bf 538} (1999) 587
[arXiv:hep-th/9703085].

\bibitem{Castro-Alvaredo:2002fc}
O.~A.~Castro-Alvaredo, A.~Fring and F.~Gohmann,
``On the absence of simultaneous reflection and transmission in 
integrable impurity systems,''
arXiv:hep-th/0201142.

\bibitem{F} P. Federbush, Phys. Rev. {\bf 121} (1961) 1247.

\bibitem{Liguori:1998xr}
A.~Liguori, M.~Mintchev and L.~Zhao,
Commun.\ Math.\ Phys.\  {\bf 194}, 569 (1998)
[arXiv:hep-th/9607085].

\bibitem{Gattobigio:1998hn}
M.~Gattobigio, A.~Liguori and M.~Mintchev,
Phys.\ Lett.\ B {\bf 428} (1998) 143
[arXiv:hep-th/9801094].

\bibitem{Gattobigio:1998si}
M.~Gattobigio, A.~Liguori and M.~Mintchev,
J.\ Math.\ Phys.\  {\bf 40} (1999) 2949
[arXiv:hep-th/9811188].

\bibitem{Mintchev:2001aq}
M.~Mintchev, E.~Ragoucy and P.~Sorba,
half  line,''
J.\ Phys.\ A {\bf 34} (2001) 8345
[arXiv:hep-th/0104079].

\bibitem{Liguori:de}
A.~Liguori, M.~Mintchev and M.~Rossi,
J.\ Math.\ Phys.\  {\bf 38} (1997) 2888.

\bibitem{MRS}
M. Mintchev, E. Ragoucy and P. Sorba, ``Boundary algebras with 
transmission'', in preparation.




\end{thebibliography}
\end{document}